\documentclass[11pt]{article}
\usepackage{graphicx}
\usepackage{amsmath,amssymb}
\def\1{{\bf 1}}

\def\be{\begin{equation}}
\def\ee{\end{equation}}

\begin{document}
\noindent {\Large \bf Molecular Dynamics Studies of Dog Prion Protein Wild-type and Its D159N Mutant}\\

\bigskip

\noindent {\Large Jiapu Zhang$^\text{ab*}$}\\
{\small {\it
\noindent $^\text{a}$Molecular Model Discovery Laboratory, Department of Chemistry \& Biotechnology,  Faculty of Science, Engineering \& Technology, Swinburne University of Technology, Hawthorn Campus, Hawthorn, Victoria 3122, Australia; 

\noindent $^\text{b}$Graduate School of Sciences, Information Technology and Engineering \& Centre of Informatics and Applied Optimisation, Faculty of Science, The Federation University Australia, Mount Helen Campus, Mount Helen, Ballarat, Victoria 3353, Australia;

\noindent $^\text{*}$Tel: +61-3-9214 5596, +61-3-5327 6335; 
jiapuzhang@swin.edu.au, j.zhang@federation.edu.au\\
}}

\noindent {\bf Abstract:} 
 {\it Prion diseases (e.g. `mad cow' disease in cattle, chronic wasting disease in deer and elk, Creutzfeldt-Jakob disease in humans) have been a major public health concern affecting humans and almost all animals. However, dogs are strongly resistant to prion diseases. Recently, through transgenic techniques, it was reported that the single (surface) residue D159 is sufficient to confer protection against protein conformational change and pathogenesis, thus provides conformational stability for dog prion protein (Neurobiology of Disease 95 (November 2016) 204-–209). This paper studies dog prion protein wild-type and D159N mutant through molecular dynamics (MD) techniques. Our MD results reveal sufficient structural informatics on the residue at position 159 to understand the mechanism underlying the resistance to prion diseases of dogs.\\ 
}


\section{Introduction}
Unlike bacteria and viruses, which are based on DNA and RNA, prions are unique as disease-causing agents since they are misfolded proteins. Prions propagate by deforming harmless, correctly folded proteins into copies of themselves. The misfolding is irreversible. Prions attack the nervous system of the organism, causing an incurable, fatal deterioration of the brain and nervous system until death occurs. Some examples of these diseases are `mad cow' disease (BSE) in cattle, chronic wasting disease (CWD) in deer and elk, and Creutzfeldt-Jakob disease (CJD) in humans.\\

Not every species is affected by prion disease. Rabbits, water buffalo, horses, and dogs are resistant to prion diseases \cite{zhang_book2015}. The research question arises: What are reasons allowing the protein of a resistant species to retain its folding? For rabbit normal cellular prion protein (PrP$^\text{C}$, where the structural region of a PrP$^\text{C}$ consists of $\beta$-strand 1 ($\beta$1), $\alpha$-helix 1 ($\alpha$1 or H1), $\beta$-strand 2 ($\beta$2), $\alpha$-helix 2 ($\alpha$2 or H2), $\alpha$-helix 3 ($\alpha$3 or H3), and the loops linking them each other), multiple amino acid residues G99, M108, S173, I214, together inhibit formation of its abnormal isoform \cite{vorberg_etal2003}. For dog PrP, D159 is the key protective residue that provides conformational stability and confers protection against prions, suggesting that a single amino acid D159 is sufficient to prevent PrP conformational change and pathogenesis \cite{sanchez-garcia_etal2016}. This paper will specially study the residue D159 of dog PrP from the protein structural dynamics point of view.\\

First, let us review some known literatures on dog PrP (and we particularly focus on its structural bioinformatics). There are a lot of literatures reporting dogs (and other canines) are rare animals being resistant to prion diseases, for examples \cite{fernandez-funez_etal2011, hasegawa_etal2013, khan_etal2010, kirkwoodc1994, nystromh2015, polymenidou_etal2008, qing_etal2014, vidal_etal2013, wopfner_etal1999, zhangw_etal2013}. 
\begin{enumerate}
\item[$-$] Early in 1994, it was found that not a single case of prion disease has been described among dogs through the exposure of dogs to prions (being fed prion-infected pet food) \cite{kirkwoodc1994}. 
\item[$-$] In 1999, it was reported that two differences between feline and canine PrP sequences, at codons 187 and 229, both involve substitutions to Arg residues which, together with the His-Arg substitution at codon 177 common to cat and dog, would increase the total positive surface charge on the molecule - this might in turn affect the potential intermolecular interactions critical for cross-species transmission of prion disease \cite{wopfner_etal1999}. 
\item[$-$] In 2004, it was reported that the three substitutions in positions 108, 164, and 182 are unique to the canine species and are thus candidates for causing a substantial species barrier, and dogs are among the few mammals that neither contain Asn at position 164 (or 159) nor His at position 182 (or 177) \cite{lysek2003, lysek_etal2004}.
\item[$-$] In 2005, the NMR structure of dog PrP was released (PDB entry 1XYK) and it was reported that the residues at positions 159 and 177 have unique charge distribution patterns on the front as well as the back side of dog PrP$^\text{C}$ \cite{lysek_etal2005}.  The residue D159 (less defined by NMR) is proposed to change the surface charge \cite{lysek_etal2005} due to its sticking out acidic side chain. 
\item[$-$] In 2006, the open reading frame of the prion protein (Prnp) gene from 16 Pekingese dogs was cloned and screened for polymorphisms \cite{wu_etal2006}. One nucleotide polymorphism (G489C) was found; the G to C nucleotide substitution results in a glutamic to aspartic acid change at codon 163; E/D163 and asparagine 107 in canine prion protein genes were replaced by asparagine and serine, respectively, in all the prion protein genes examined \cite{wu_etal2006}.
\item[$-$] In 2008, transmission experiments in Madin Darby canine kidney (MDCK) cells showed they do not replicate human CJD prions and mouse (infected with scrapie) prions \cite{polymenidou_etal2008}, supporting the resistance of dogs to prions. Human PrP$^{\text{C}}$ is selectively targeted to the apical side of the MDCK \cite{de_keukeleire_etal2007}.
\item[$-$] In 2009, Onizuka (2009) reported the substitutions N104G and S107N have the biggest impact to the conformational transition and stability of dog PrP \cite{onizuka2009}. In 2009,  it was reported in \cite{wan_etal2009} that the three substitutions in positions 107, 163, and 181 are unique to the Arctic fox and dog, and these substitutes might be associated with susceptibility and species barriers in prion diseases.
\item[$-$] In 2010, Khan et al. (2010) found that at pH 4.0 dogs have the lowest concentration of $\beta$ intermediate state compared with hamsters, mice, rabbits and horses \cite{khan_etal2010, qing_etal2014} -– this can be adopted in evaluating the prion disease susceptibility of each species (hamster$>$mouse$>$rabbit$>$horse$>$dog) \cite{ fernandez-funez_etal2011, khan_etal2010, qing_etal2014}.
\item[$-$] In 2013, Hasegawa et al. (2013) reported that there are large differences in local structural stabilities between canine and bovine PrP, and this appearance might link diversity in susceptibility to BSE prion infection \cite{hasegawa_etal2013}. 
\item[$-$] In 2015, a survey of prnp genes implied that the prion disease resistant canines harbor amno acids DRK in positions 159, 177, and 185 \cite{nystromh2015}. Arg177 in H2 of dog PrP also causes unique charge distribution patterns on the front and the back side of dog PrP$^\text{C}$ \cite{lysek_etal2005}, but Arg177 are positively charged residue that should have minor effects on PrP structure and surface charge \cite{sanchez-garcia_etal2016}, because the sequence of the rigid $\beta$2-$\alpha$2 loop in the dog is identical to mouse PrP, indicating that this region does not contribute to the conformational stability of dog PrP \cite{fernandez-funez_etal2011}. 
\end{enumerate}
\noindent In 2016, Sanchez-Garcia et al. (2016) reported that a single amino acid (D159) from the dog PrP suppresses the toxicity of the mouse PrP in Drosophila \cite{sanchez-garcia_etal2016} and their laboratory experience suggested that a single D159 substitution is sufficient to prevent PrP conformational change and pathogenesis \cite{sanchez-garcia_etal2016}. D159 is a unique amino acid found in PrP from dogs and other canines that was shown to alter surface charge. The acidic amino acid D159 is on the $\alpha$1-$\beta$2 loop and exposed on the surface of dog PrP, resulting in increased negative charge \cite{sanchez-garcia_etal2016}. The $\beta$1-$\alpha$1 and $\alpha$1-$\beta$2 loops interact more closely in dog PrP than in susceptible animals thus the subtle changes in the orientation of the side chains and the closer loops may affecting the stability of the $\beta$-sheet \cite{hasegawa_etal2013, lysek_etal2005} -– this might explain the poor NMR resolution of residue D159. The mutation N159 will create a neutral surface that extends over the surface of the two loops (i.e. the $\beta$1-$\alpha$1 loop and the $\alpha$1-$\beta$2 loop) and H2 \cite{sanchez-garcia_etal2016}. In the D159 region there only harbors one nonsense pathogenic mutation Q160X, however, this critical domain should be investigated in more details, because the identification of $\alpha$1-$\beta$2 loop-binding proteins are expected to reveal clues about the molecular mechanisms and the extrinsic factors mediating PrP conversion from soluble normal prion protein PrP$^\text{C}$ (predominant in $\alpha$-helices) to insoluble diseased infectious prions PrP$^\text{Sc}$ (rich in $\beta$-sheets) \cite{sanchez-garcia_etal2016}. It has been proposed that this change in surface charge will result in altered interactions with other proteins, possibly proteins that contribute to PrP conversion \cite{sanchez-garcia_etal2016}. The $\alpha$1-$\beta$2 loop is highly conserved among mammals but only dog PrP possesses the unique acidic residue D159 suggesting that D159 plays a role in providing {\sl global} stability to dog PrP \cite{fernandez-funez_etal2011}. We also once reported that the residue at position 159 is unique in dog PrP$^\text{C}$ \cite{zhang2012, zhang_book2015, zhangl2011}. This paper will continue our research on the molecular dynamics (MD) studies of dog PrP, especially on the MD studies of its D159N mutant and their comparisons.\\

The rest of this paper is organized as follows. In Section 2 we will give the MD simulation materials and methods. Section 3 will presents the MD results and their analyses (where surface charge distributions are specially focused to analyze the MD trajectories). After the Results and Discussions section, some Concluding Remarks (revealed from the MD to understand the mechanism underlying the resistance to prion diseases of dogs) will be given in the last section of this paper.

\section{Materials and Methods}
The MD simulation materials and methods are completely as the ones of \cite{zhang2012, zhang_book2015, zhangl2011}. The D159N mutant model used in this study was constructed by one mutation D159N at position 159 using the NMR structure 1XYK.pdb of dog PrP (121--231), where the NMR experimental temperature is 293 K (i.e. the room temperature), pH value is 4.5, and pressure is AMBIENT. To neutralize the MD systems by sodium ions, 2 Na+ ions were added to the wild-type, and 1 Na+ ion was added to the D159N mutant (because the residue Asn is without charge). 1XYK.pdb has 8 Arg+ residues and 4 Lys+ residues (and 2 His+ residues), 6 Asp- residues and 8 Glu- residues, two salt bridges D144.OD1--R+148.NH1/2, D178.OD2--R+164.NE/NH2 and one cation-$\pi$ interaction R164.CD-Y128 detected by FirstGlance in Jmol (bioinformatics.org/firstglance/fgij/). This paper uses Maestro 9.7 2014-1 (Academic use only) free package to draw the Poisson–Boltzmann electrostatic potential surface charges of the external polarization: we choose indi = 1.0 as the solute dielectric constant, exdi = 80.0 as the solvent dielectric constant, 12 $\textrm{\AA}$ as the solvent radius, and 300 K as the Temperature: EPS mapped on molecular surface is chosen.\\

Electrostatic potential surfaces are valuable in structure-based/computer-aided drug design because they help in optimization of electrostatic interactions between the protein and the ligand. These surfaces can be used to compare different inhibitors with substrates or transition states of the reaction.  To study the surface charge of a residue and its local and global impacts should firstly consider the salt bridges (SBs) it linked with \cite{zhang_etal2015}. SBs are calculated by oppositely charged atoms that are within 6.5 $\textrm{\AA}$ and are not directly hydrogen-bonded. It should be the average charge calculated per residue or the specific atom charge of the residue. The donor residues involved are Asp-, Glu-, and the acceptor residues involved are Lys+, Arg+, His+, and the real computed distance is within 6.5 $\textrm{\AA}$ in Amber package.

\section{Results and Discussions}
\begin{figure}[h!] 
\centerline{
\includegraphics[width=5.2in]{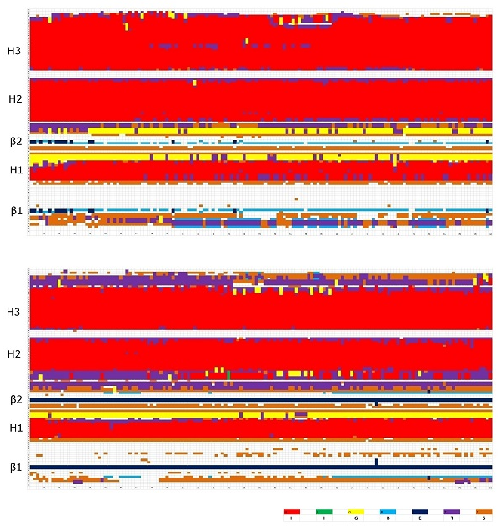}
}
\caption{\textsf{{\small {\sl Secondary Structure graph for dog PrP D159N mutant and dog PrP wild-type (from up to down) at 300 K under neutral pH environment (x-axis: time (0–30 ns), y-axis: residue number; red: $\alpha$-helix (H), blue: $\beta$-sheet (E), where H = $\alpha$-helix, B = residue in isolated $\beta$-bridge, E = extended strand, participates in $\beta$-ladder, G = 3-helix (3$_\text{10}$ helix), I = 5 helix ($\pi$-helix), T = hydrogen bonded turn, and S = bend).}}}}  \label{2ndStruct_d159-dog_300K-pH7-712}
\end{figure}

Firstly, we see the secondary structure changes of the D159N mutant and of the wild-type during the whole 30 ns' molecular movement. By Fig. \ref{2ndStruct_d159-dog_300K-pH7-712}, we may see what we want: for $\beta$1 and $\beta$2, the wild-type has the clear extended $\beta$-strand (participates in $\beta$-ladder) structure (with the occupied rate 3.66\% during the whole 30 ns), but the D159N mutant has changed into $\beta$-bridge structures (occupied rate 0.34\% during 30 ns). This performance implied to us the mutation D159N has clearly changed the PrP structure in domains of $\beta$1, the $\beta$1-$\alpha$1 loop, H1, the $\alpha$1-$\beta$2 loop, $\beta$2, and the $\beta$2-$\alpha$2 loop. The mutation made the stable wild-type structure (before H2) become unstable.

\begin{figure}[h!] 
\centerline{
\includegraphics[width=2.6in]{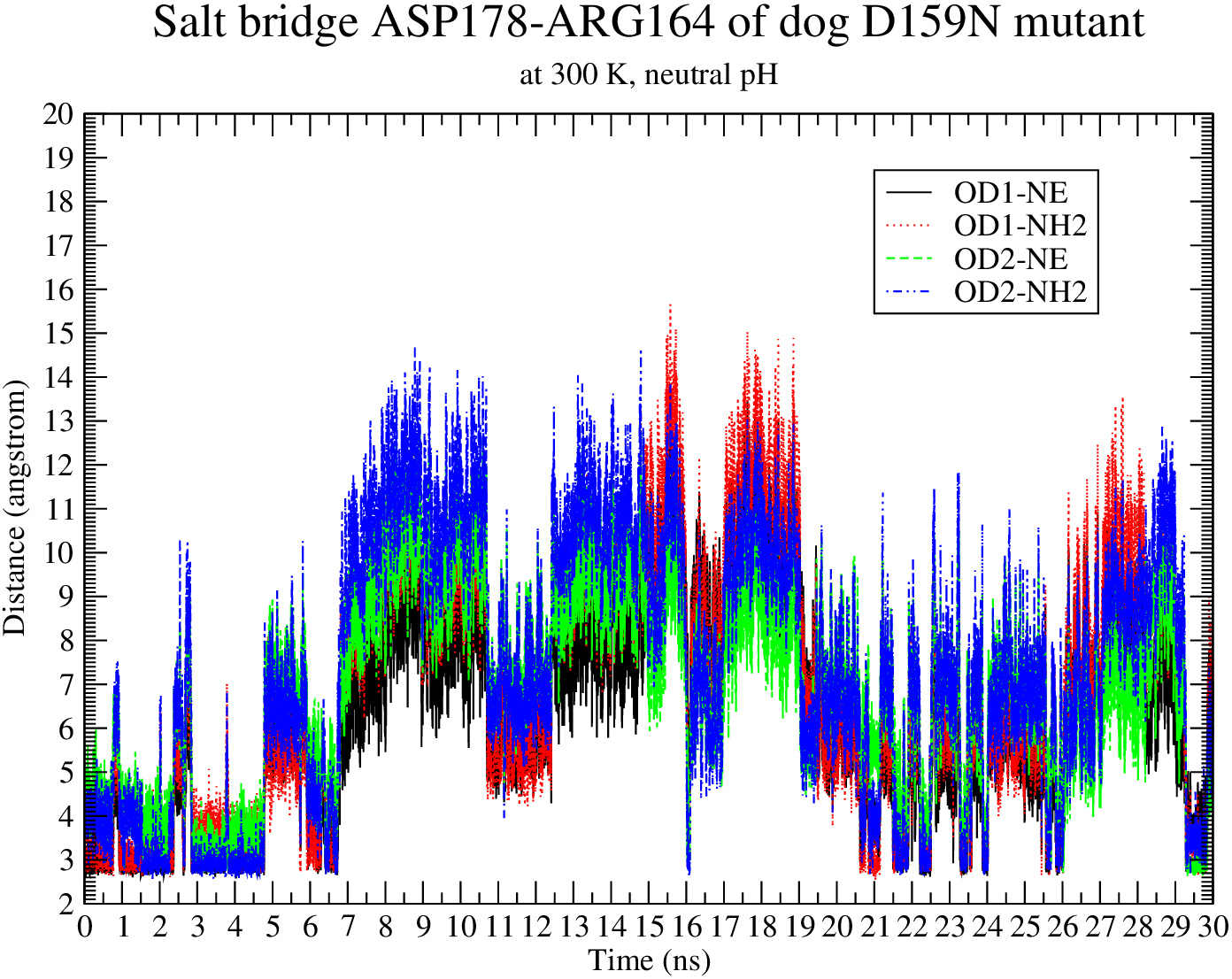} \qquad
\includegraphics[width=2.6in]{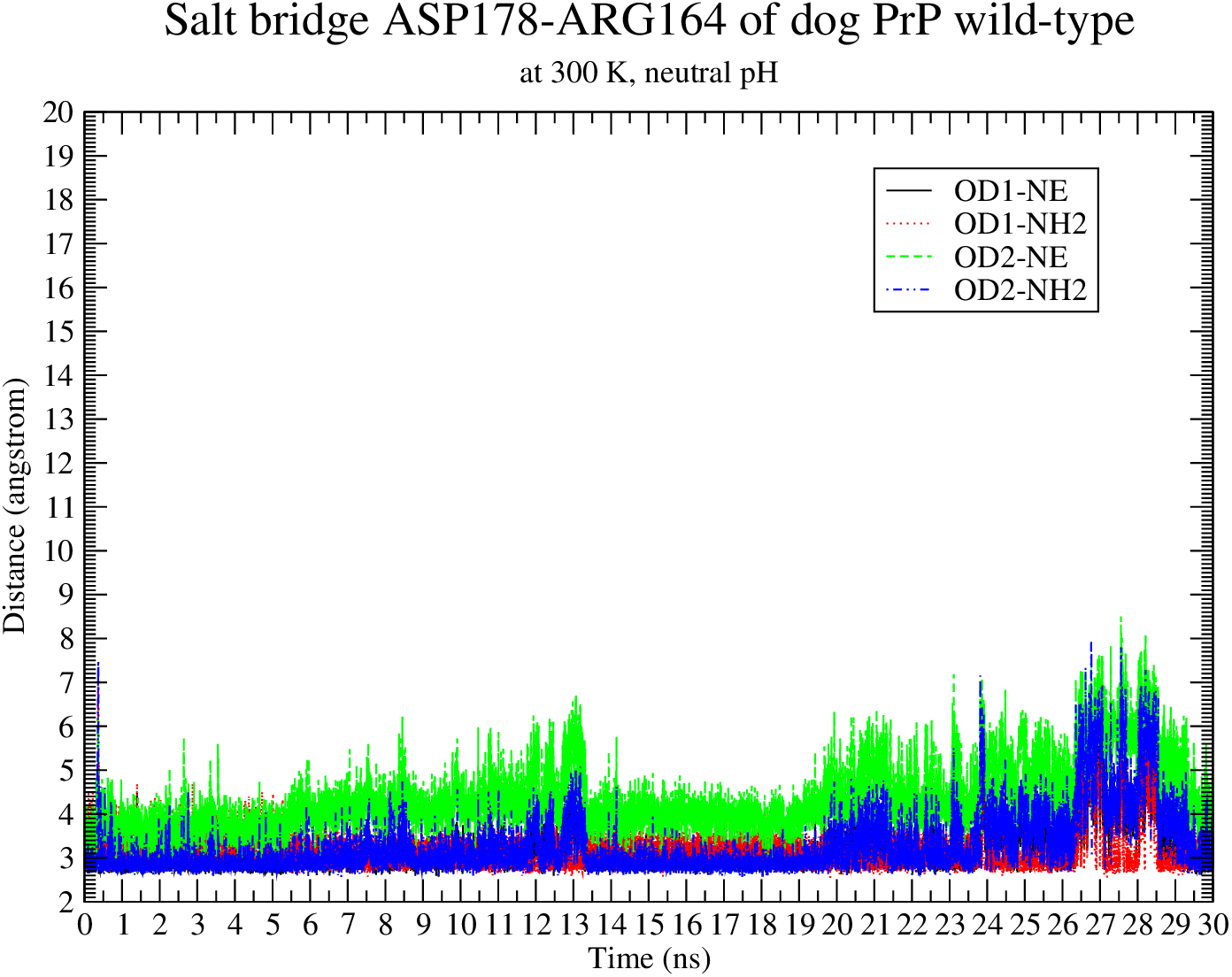}
}
\caption{\textsf{{\small {\sl The SB D178--R164 of the D159N mutant (with occupied rate 9.25\% during the whole 30 ns of MD) and of the wild-type (with occupied rate 49.43\% during the whole 30 ns of MD) at 300 K under neutral pH condition..}}}}  \label{SB_Asp178-Arg164-d159-dog_300K-pH7-712}
\end{figure}

\begin{figure}[h!] 
\centerline{
\includegraphics[width=5.2in]{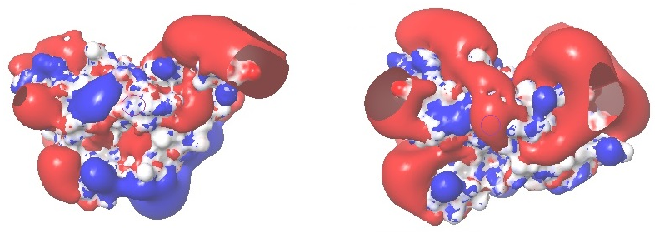}
}
\caption{\textsf{{\small {\sl Surface charge distributions of the 30 ns' average structure. Left: The N159 mutant of dog PrP; right: The wild-type of dog PrP. The circles indicate charge differences originating at position 159. Red, blue, and white indicate the negative, positive charges and uncharged respectively.}}}}  \label{surface-charge_d159-dog_300K-pH7-712-average-30ns}
\end{figure}

Secondly, we see the SBs of the D159N mutant and of the wild-type during the 30 ns' MD simulations. As we previously reported \cite{zhang2012, zhang_book2015, zhangl2011} the SB D178--R164 like a taut bow string of the $\beta$2-$\alpha$2 loop was broken by the D159N mutation (Fig. \ref{SB_Asp178-Arg164-d159-dog_300K-pH7-712}). This implies to us that residue R164 spans only 4 residues from residue D159, and residue D159 has a rather impact on the local structure. The SB D159--R136 at residue D159, with occupied rate 92.48\% during 30 ns, was broken in the D159N mutant; this SB keeps the $\beta$1-$\alpha$1 loop apart from the $\alpha$1-$\beta$2 loop. This shows the stabilizing effect of D159 in dog PrP to affect the both locally and globally unusual charge distribution of NMR structure of dog PrP. 
\noindent Furthermore, we want to see and analyze all the SBs during the whole 30 ns (Tab. \ref{d159-dogPrP_SBs_300K-pH7-712}).  
\begin{table}[h!]
\caption{\textsf{{\small {\sl All SBs of the D159N mutant and of the wild-type at 300 K, neutral pH value during the whole 30 ns' MD simulations:}}}}
\centering
{\scriptsize
\begin{tabular}{|l                     ||r                      |r        |r          ||l|} \hline 
                Salt Bridges           &D159N mutant            &         &Wild-type  &In PrP,\\
                (SBs)                  &(\% occupied rate)      &         &(\% occupied rate) &where?\\ \hline		
                D147--R+148	           &100.00	                &         &100.00     &in H1\\ \hline
                D178--R+177	            &99.82	                &         &99.57      &in H2\\ \hline
                E223--K+220	            &98.56	                &         &41.85      &in H3\\ \hline
                E207--K+204	            &98.43                  &         &96.13      &in H3\\ \hline
                E211--R+208	            &97.71                  &       * &99.13      &in H3\\ \hline
                D144--R+148	            &93.72	                &         &53.47      &in H1\\ \hline
                D147--H+140	            &82.21	                &         &72.37      &H1 -- $\beta$1-$\alpha$1-loop\\ \hline
                E207--R+208	            &76.43	                &         &44.87      &in H3\\ \hline
                E196--K+194	            &75.98                  &     	  &59.57      &in $\alpha$2-$\alpha$3-loop\\ \hline
                E152--R+148	            &53.16                  &     $<$ &53.52      &in H1\\ \hline
                E152--R+151	            &51.11	                &         &43.63      &in H1\\ \hline
                H187--R+156 	        &42.75                  &     $<$ &44.74      &H2 -- $\alpha$1-$\beta$2-loop\\ \hline
                E211--R+177	            &30.02                  &     $<$ &45.91      &H3 -- H2\\ \hline
                E221--K+220	            &14.07                  &       * &53.71      &in H3\\ \hline
                D147--R+151	            &11.41                  &     $<$ &25.68      &in H1\\ \hline
                D202--H+187	             &9.25	                &          &1.19      &H3 -- H2\\ \hline
                D178--R+164	             &7.82                  &     $<$ &49.43      &H2 -- $\beta$2-$\alpha$2-loop\\ \hline
                D167--R+228	             &4.22                  &         &           &$\beta$2-$\alpha$2-loop -- H3\\ \hline	
                E146--K+204	             &2.96                  &     $<$ &24.73      &H1 -- H3\\ \hline
                H187--K+185 	         &2.31                  &          &0.46      &in H2\\ \hline
                E196--R+156	             &0.43                  &     $<$ &16.20      &$\alpha$2-$\alpha$3-loop -- $\alpha$1-$\beta$2-loop\\ \hline
                H140--R+151 	         &0.20                  &       *  &0.29      &$\beta$1-$\alpha$1-loop -- H1\\ \hline
                E200--K+204	             &0.03                  &       *  &0.21      &in H3\\ \hline
                E196--H+187	             &0.03                  &       *  &0.09      &$\alpha$2-$\alpha$3-loop -- H2\\ \hline
                E221--R+228	             &0.02                  &          &          &in H3\\ \hline
                D159--R+136	             &                      &     $<$ &92.48      &$\alpha$1-$\beta$2-loop -- $\beta$1-$\alpha$1-loop\\ \hline
                D202--R+156	             &                      &     $<$  &8.20      &H3 -- $\alpha$1-$\beta$2-loop\\ \hline
                E223--R+228	             &                      &       *  &0.15      &in H3\\ \hline
                D144--H+140	             &                      &       *  &0.05      &in $\beta$1-$\alpha$1-loop\\ \hline
                H140--R+228	             &                      &       *  &0.01      &$\beta$1-$\alpha$1-loop -- H3\\ \hline
\end{tabular}
} 
\label{d159-dogPrP_SBs_300K-pH7-712}
\end{table}
\noindent Seeing Table \ref{d159-dogPrP_SBs_300K-pH7-712}, we may know that ``$<$" shows the local impact of the D159N mutation which made the weaker of the wild-type's SBs such as D159--R+136, E211--R+177, D147--R+151, D178--R+164, E146--K+204, E196--R+156,  D202--R+156, H+187--R+156, E152--R+148, and ``*" implies the global impact of the D159N mutation which made weaker of the wild-type's SBs. To understand better the above SBs, we illuminate the surface charge distributions of the 30 ns' average structure of the D159N mutant and the wild-type respectively (Fig. \ref{surface-charge_d159-dog_300K-pH7-712-average-30ns}).
\noindent From Fig. \ref{surface-charge_d159-dog_300K-pH7-712-average-30ns}, we see that the D159N mutation made the negative charges and the positive charges redistributed around the residue at position 159; however, the negative charges covering H1 and the tail of H3 are not changed very much - this implies to us we had better not seek drug target(s) from H1 or the tail of H3.\\

Thirdly, we will give a brief overview of the changes of some hydrogen bonds (HBs) and hydrophobic interactions (HYDs) made by the D159N mutation. In view of the number of HBs, we cannot say differences between the D159N mutant and the wild-type; however, the HBs listed in Tab. \ref{d159-dogPrP_HBs_300K-pH7-712} contribute to the structural stability of wild-type dog PrP more than of the D159N mutant. We may see from Tabs. \ref{d159-dogPrP_SBs_300K-pH7-712}$\sim$\ref{d159-dogPrP_HBs_300K-pH7-712} that (i) D202--R+156 and E223--R+228 are two polar contacts for the wild-type, but the D159N mutant is without these polar contacts; (ii) D147--H+140, E211--R+177, D147--R+151, D178--R+164 are four strong polar contacts for the wild-type but weaker for the D159N mutant.
\begin{table}[h!]
\caption{\textsf{{\small {\sl All HBs of the D159N mutant and of the wild-type at 300 K, under neutral pH environment during the whole 30 ns' MD simulations:}}}}
\centering
{\scriptsize
\begin{tabular}{|l                     ||r                 |r|} \hline 
                Hydrogen bonds         &D159N mutant       &Wild-type\\
                (HBs)                  &(\% occupied rate) &(\% occupied rate)\\ \hline	
T188@O--T192@HG1     & 89.69  &90.84\\ \hline
C179@O--Y183@HG1     & 86.21  &89.55\\ \hline
D178@OD1--R+164@HE   & 13.13  &54.05\\
D178@OD2--R+164@HH21 &  9.18  &52.07\\
D178@OD1--R+164@HH21 &  6.95  &38.84\\ \hline
D202@OD2--R+156@HH21 &        &46.46\\
D202@OD1--R+156@HH21 &        &38.87\\
D202@OD1--R+156@HE   &        &37.73\\
D202@OD2--R+156@HE   &        &17.25\\ \hline
D202@OD1--Y157@HH    &  9.74  &53.98\\ \hline
D202@OD2--T199@HG1   &        &41.63\\
D202@OD1--T199@HG1   &        &13.73\\ \hline
D202@OD2--Y149@HH    &        &24.72\\ 
D202@OD1--Y149@HH    &        &17.85\\ \hline
H+187@O--T191@HG1    & 71.25  &84.35\\ \hline
Y149@O--N153@HD21    &        &17.46\\ \hline
Q212@O--T216@HG1     & 48.43  &81.55\\ \hline
Q227@OE1--Q212@HE21  &        &13.29\\
Q212@OE1--Q227@HE21  &         &9.74\\ \hline
D147@OD1--R+151@HH12 &  7.00  &26.35\\
D147@OD1--R+151@HH11 &  6.75  &24.47\\ \hline
D147@OD2--H+140@HD1  & 16.73  &26.46\\ \hline
D144@OD1--R+151@HH12 &        &11.39\\ \hline
V189@O--T193@HG1     & 21.28  &32.71\\ \hline
E211@OE2--R+177@HH21 & 11.91  &15.17\\
E211@OE1--R+177@HH21 &  8.41  &18.94\\
E211@OE2--R+177@HE   &        &29.98\\
E211@OE1--R+177@HE   &        &26.05\\
E211@OE1--R+177@HH12 &         &8.71\\ \hline
E207@OE2--R+177@HH22 &  6.55  &10.69\\
E207@OE1--R+177@HH22 &  5.21  &14.15\\
E207@OE1--R+177@HH12 &         &6.57\\ \hline
E223@O--R+228@HH21   &        &10.76\\
E223@O--R+228@HE     &        &11.24\\
E223@O--R+228@HH11   &         &7.45\\ \hline
E223@O--Q219@HE22    &        &13.09\\ \hline
E146@OE2--K+204@HZ2  &  6.34  &18.46\\
E146@OE2--K+204@HZ1  &  5.96  &16.90\\
E146@OE1--K+204@HZ2  &  5.94  &17.65\\
E146@OE2--K+204@HZ3  &  5.69  &15.90\\
E146@OE1--K+204@HZ3  &  5.54  &15.75\\
E146@OE1--K+204@HZ1  &  5.31  &16.76\\ \hline
Q219@OE1--Q227@HE22  &        &24.37\\
Q227@OE1--Q219@HE22  &         &5.71\\ \hline
T216@OG1--Q227@HE21  &         &6.78\\ \hline
S231@OG--N171@HD22   &         &9.12\\ \hline
G229@O--N171@HD21    &        &28.10\\ \hline
H+140@O--R+208@HH21  &        &22.37\\
H+140@O--R+208@HE    &        &11.59\\ \hline
P158@O--R+136@HH12   &        &15.63\\ \hline
V176@O--Y218@HH      &        &14.65\\ \hline
D178@OD1--Y128@HH    &         &8.95\\ \hline
Q186@OE1--Y128@HH    &         &8.62\\ \hline
S132@OG--Q217@HE21   &         &7.61\\ \hline
N174@OD1--Q172@HE22  &         &7.36\\ \hline
A224@O--R+228@HH11   &         &6.95\\ \hline
Y225@O--R+228@HE     &         &6.63\\ \hline
N197@O--R+156@HH22   &         &6.33\\ \hline
E196@OE2--R+156@HH12 &         &5.33\\ \hline
\end{tabular}
} 
\label{d159-dogPrP_HBs_300K-pH7-712}
\end{table}
\noindent Regarding hydrophobic interactions (HYDs), throughout the whole 30 ns' MD simulations, the D159N mutation has a local impact - it made HYD M213--V161 become weaker than in the wild-type, and has global impacts - it made V215--M213 and V209--I205 become weaker than in the wild-type. Around the residue at position 159, we find there are one $\pi$-$\pi$ stacking F141--Y150 and one $\pi$-cation Y164--R164 in the D159N mutant and the $\pi$-$\pi$ stacking F141--Y150 in the wild-type PrP through the whole 30 ns' protein movement. We also noticed that GN8 \cite{kuwata_etal2007, hosokawa-muto_etal2012}, an antiprion drug 
fixing the distance between N159 and E196 being 1.54 $\textrm{\AA}$, was designed at the position 159 - this might show the importance of the PrP residue at position 159. 

\section{Concluding Remarks}
The mutation D159N of dog PrP had profound effects on protein structure of dog PrP: (1) it altered the surface charge distribution, both locally and globally, (2) it reduced the stability of the tertiary structure of dog PrP, and (3) it increased the mobility of dog PrP structure before H2. The MD studies of this paper confirmed these three findings and emphasized the contribution of the single residue D159 in dictating the global (and local) charge distribution and structural stability of dog PrP. This paper presented detailed sufficient structural informatics on the residue at position 159 to understand the mechanism underlying the resistance to prion diseases of dogs; this may be useful for the medicinal treatment of prion diseases. 
 
\section*{Acknowledgments}
This research was supported by a Victorian Life Sciences Computation Initiative (VLSCI) grant numbered FED0001 on its Peak Computing Facility at the University of Melbourne, an initiative of the Victorian Government (Australia).

\end{document}